\documentclass[aps,psfig,pre,preprint,groupedaddress]{revtex4}
\usepackage{graphicx}
\usepackage{amsmath}

\begin{document}


\title{Coarse-graining the Dynamics of a Driven Interface in the Presence of Mobile Impurities:
Effective Description via Diffusion Maps}


\author{Benjamin E. Sonday}
\email[]{bsonday@math.princeton.edu}
\affiliation{Program in Applied and Computational Mathematics,
Princeton University, Princeton, NJ 08544}
\author{Mikko Haataja}
\email[]{mhaataja@princeton.edu}
\homepage[]{http://www.princeton.edu/~mhaataja/}
\affiliation{Princeton Institute for the Science and Technology of
Materials (PRISM) and Department of Mechanical and Aerospace
Engineering, Princeton University, Princeton, NJ 08544}
\author{Ioannis G. Kevrekidis}
\email[]{yannis@princeton.edu}
\homepage[]{http://arnold.princeton.edu/~yannis/}
\affiliation{Department of Chemical Engineering and Program in
Applied and Computational Mathematics, Princeton University,
Princeton, NJ 08544}


\date{\today}

\begin{abstract}

Developing effective descriptions of the microscopic dynamics of
many physical phenomena can both dramatically enhance their
computational exploration and lead to a more fundamental
understanding of the underlying physics.
Previously, an effective description of a driven interface in the
presence of mobile impurities \cite{mikko}, based on an Ising
variant model \cite{mend2001} and a single empirical coarse
variable, was partially successful; yet it underlined the necessity
of selecting additional coarse variables in certain parameter
regimes.
In this paper we use a data mining approach to help identify the
coarse variables required.
We discuss the implementation of this \textit{diffusion map}
approach, the selection of a similarity measure between system
snapshots required in the approach, and the correspondence between
empirically selected and automatically detected coarse variables.
We conclude by illustrating the use of the diffusion map variables
in assisting the atomistic simulations, and we discuss the
translation of information between fine and coarse descriptions
using lifting and restriction operators.

\end{abstract}


\maketitle

\section{Introduction\label{Intro}}
Driving an extended defect or domain wall in a system that contains
stationary or mobile impurities lies at the heart of many physical
phenomena, ranging from the classic example of the motion of grain
boundaries in polycrystalline materials \cite{cahn62} to the more
recent modeling of ferroelectric domain wall dynamics
\cite{tybell2002}, dislocation dynamics in metals containing
stationary and mobile impurities \cite{hahner}, and stick-slip
phenomena in tribology \cite{cule96}.
While the interactions between isolated impurities and the domain
wall are rather well understood at the microscopic level,
translating the impact of such interactions to effective dynamics at
mesoscopic or macroscopic scales remains a challenging problem.

In this work, we employ a microscopic model, a variant of the Ising
model in which the domain wall moves under the influence of an
external magnetic field while the mobile interstitial impurities are
attracted to the domain wall.
Our goal is to extract an effective description of the system based
on a small number of coarse variables.
In previous work \cite{mikko}, a one-dimensional description was
proposed based on empirical evidence and physical intuition.
While this reduced description accurately represents the Ising
variant model in some regimes, it fails to capture the more salient
features of the coupled impurity-domain wall dynamics.
Clearly, a more systematic dimensionality reduction (i.e.,
coarse-graining) tool, which relies less on intuition, is desired.

To this end, diffusion maps have recently emerged as an effective
nonlinear dimensionality reduction tool
\cite{belkin2003,coifman1,coifman2}, see also \cite{tenen2000}.
Diffusion maps find a ``best'' (possibly nonlinear) low-dimensional
underlying manifold that (approximately) contains the data set of
interest.
Given a data set of system states, the diffusion map approach first
requires a scalar \textit{similarity measure} between each pair of
system states.
When the similarity measure is based upon dynamic proximity, it is
expected that the resulting manifold parameterizes the meaningful
dynamics of the system states provided.
When the data are in the form of vectors with real entries (e.g.
concentrations of chemical species in a chemical reaction network)
the standard Euclidean distance may be used to construct such a
similarity measure (presumably, system states with similar
concentration profiles are dynamically close as well).
Typically (see the Appendix) the similarity measure becomes
negligible beyond a local neighborhood of each data vector: large
Euclidean distances do not provide useful similarity information.
This is related to the notion that on a general nonlinear manifold,
large Euclidean distances can not be expected to reliably
approximate geodesic distances.
These pairwise similarities are used in the diffusion map algorithm
to construct a matrix (the discretization of a diffusion operator)
whose leading eigenvectors (discretizations of the corresponding
eigenfunctions) are computed.
If a spectral gap is detected between a few leading eigenvalues and
the remaining ones, the corresponding few eigenvectors can help
\textit{nonlinearly} embed the data set in $\mathbf{R}^n$, thereby
parameterizing the underlying manifold.
The Euclidean distance between snapshots in the new, reduced space,
is the \textit{diffusion distance}, a notion similar to geodesic
distance on a manifold (see, however, \cite{lafonlee}); the process
thus transforms our local understanding of similarity between data
to a global geometry.

When this data-mining approach is applied to the Ising variant model
simulations, we find that we can extend the previously obtained
one-dimensional coarse variable description to a more quantitative
description involving two coarse variables.
Treating the Ising model variant as a stochastic dynamical system in
these two coarse variables, we postulate a two-dimensional Langevin
(and the corresponding Fokker-Planck) equation description.
The effect of all other degrees of freedom is incorporated in the
drift and diffusion coefficients of the Fokker-Planck equation; we
extract these coefficients through either long-term or multiple
short-burst Monte Carlo simulations, and use them to extract
system-level statistics for the problem.

Dimensionality reduction of physical systems and their models can be
useful for many reasons.
An effective reduced description provides intuition about the
physical system, grants accelerated retrieval of dynamical
quantities of interest, and allows for increased computational
efficiency.
The Ising model variant discussed in this paper has high underlying
dimensionality and no transparent effective description.
Even when there is no clear correspondence between diffusion map
coordinates and physical variables, the computer-assisted detection
of coarse variables can still be a useful step in gaining intuition
about which physical features of the model are dynamically
important.

The rest of this paper is organized as follows.
Key features of the Ising variant model and the dynamics it exhibits
are presented in section \ref{keyfeat}.
The choice of a similarity measure between nearby system snapshots
and its use in the data-mining process is described in section
\ref{distmet}.
The extraction of a two-dimensional effective stochastic description
follows in section \ref{resultdescr}.
The selection of a physically meaningful second coarse variable is
also explored here.
Finally, the translation of information between the coarse and the
physical system descriptions is discussed in section \ref{lifting},
where we also show how to accelerate the calculation of dynamical
quantities of interest using our effective description.
The diffusion map algorithm is discussed in the Appendix.

%


\section{The kinetic Monte Carlo simulation and its dynamics\label{keyfeat}}

Our kinetic Monte Carlo simulations are performed on a $32\times32$
square lattice (grid of domain spins) with associated interstitial
sites, each of which may or may not contain an impurity; this is the
same model simulated in \cite{mikko}, and we describe it here only
briefly for completeness.
Evolution on this lattice is governed by the Hamiltonian
\begin{equation*}
\mathcal{H} / k_BT = -J/2 \sum s_i s_j - h \sum s_i + E_0/4 \sum
\epsilon_\alpha |\sum s_j|,
\end{equation*}
where $h$ is the external magnetic field, acting normal to the
lattice plane.
Individual spins ($``+1"$ or $``-1"$) are denoted as $s_i$, $J$
accounts for the domain wall energy, $E_0 > 0$ quantifies the
attraction of the impurities by the domain wall, and
$\epsilon_\alpha = 1$ if an impurity lies in interstitial site
$\alpha$ and $0$ otherwise; $c_{imp}$ is the impurity concentration.

\begin{figure}
\includegraphics[height=170mm,width=130mm,keepaspectratio=false,angle=270]{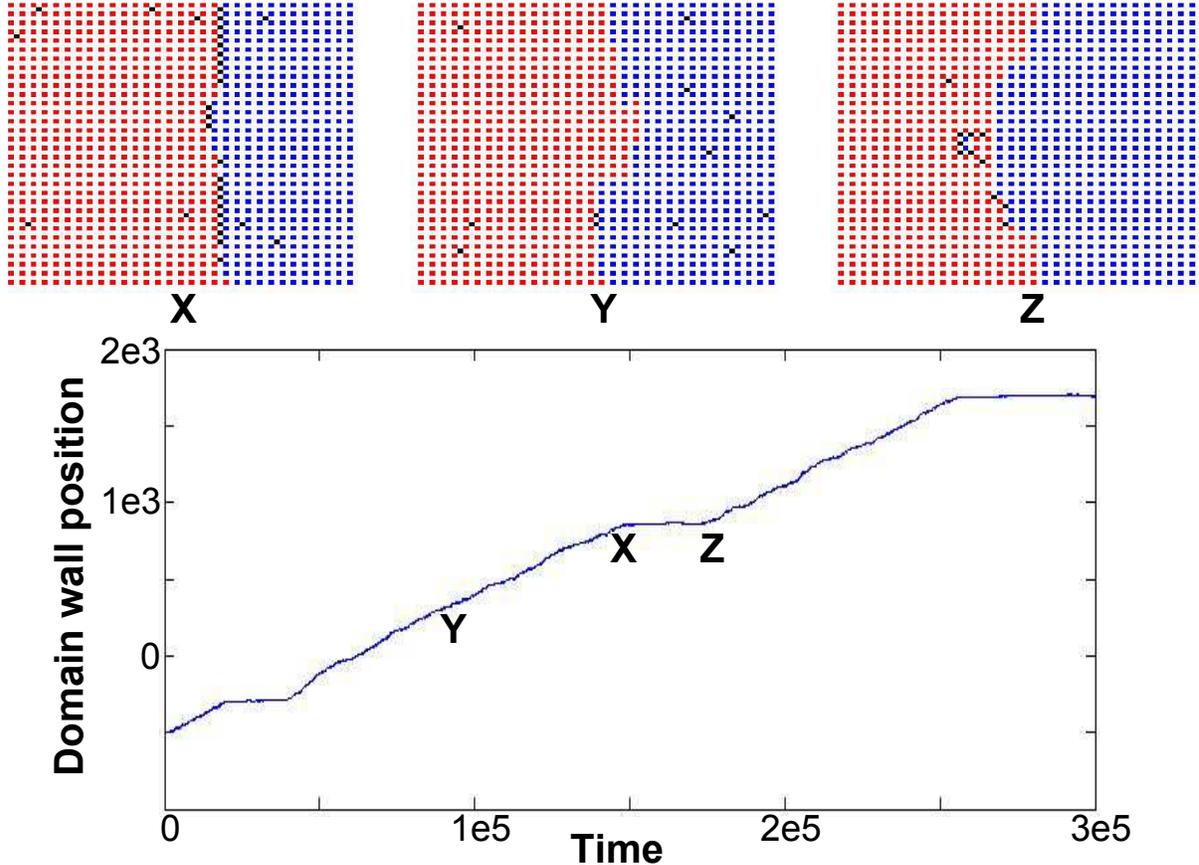}
\caption{Three typical system snapshots (above) arising in a
sequence of ``stick-slip" events.
Impurities are black dots, $+1$ spins are red, and $-1$ spins are
blue; the domain extends infinitely to the left(right) of the the
part shown and it is periodic in the vertical direction.
The domain wall in snapshot X is practically stationary; in snapshot
Y it is moving (more or less) smoothly to the right; in snapshot Z
it just started moving smoothly to the right after ``engulfing" its
domain wall impurities (see text).\label{typsiss}}
\end{figure}

The domain is periodic in the vertical direction and extends, in
principle, infinitely far to the left and right \cite{mikko}:  all
domain spins to the left of the simulation grid are assumed to be
$+1$ and all domain spins to the right to be $-1$.
We can think of the system as lying on an infinitely long cylinder
of circumference $32$.
Our simulations always involve a \textit{domain wall} (a boundary
between $+1$ spins and $-1$ spins), and our chosen cylinder
\textit{length} of $32$ is sufficient to contain the extent of this
domain wall.
Three typical system snapshots are shown in Fig.\ref{typsiss} (in
our figures we will typically ``crop" the domain in length, but not
circumference, so that only the portion containing the domain wall
is shown).
As the simulation evolves, the domain wall moves to the right under
the influence of the external magnetic field; to ``keep up" with
this motion, we continually shift the computation along the axis of
the cylinder, keeping the domain wall close to the middle of our $32
\times 32$ simulation grid.
The impurities diffuse interstitially and are attracted to the
domain wall \cite{mend2001}; the boundary conditions for impurity
motion are periodic in both the $x$ and the $y$ directions.
At each time step of the kinetic Monte Carlo simulation, we select
either a domain spin to flip or an impurity to move.
The resulting energy change is calculated, and the move is accepted
if the energy change is negative, or with probability
$e^{-\frac{\Delta \mathcal{H}}{k_BT}}$ if it is positive.
The simulation clock advances by $\frac{1}{32 \times 32}$ units
whenever a spin is sampled.
In our simulations, we used the parameters $h= 0.12$, $c_{imp} =
0.01$, $J = 2.0$, and $E_0 = 5.75$.

\begin{figure}
\includegraphics[height=170mm,width=92mm,keepaspectratio=false,angle=270]{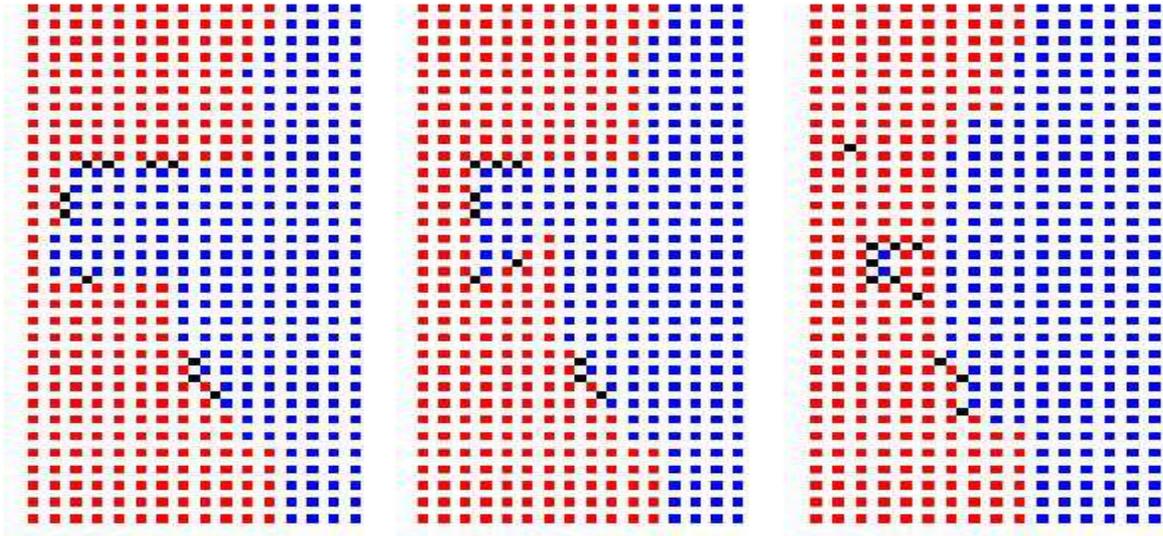}
\caption{Three system snapshots just before (left), during (middle),
and just after (right) a transition from trapped domain wall to
smoothly moving domain wall.
Transitions like this one suggest that the number of impurities on
the domain wall may not completely describe the system state, and
that domain wall shape may also become important.\label{rarEvent}}
\end{figure}

Diffusing impurities impede domain wall movement, while the external
magnetic field promotes domain wall motion.
Large external fields and small impurity-wall attraction result in
smoothly traveling domain walls, while small external fields and
large impurity-wall attraction result in walls which hardly move.
The parameters in Fig.\ref{typsiss} correspond to a compromise
between these two competing forces.
As the snapshots in Fig.\ref{typsiss} summarize, the model is
capable of exhibiting a type of bistability: it switches between a
(more or less) constant speed motion to the right and (more or less)
stationarity.
One metastable state corresponds to an impurity-rich, stationary
domain wall, while the second metastable state corresponds to an
impurity-deficient, mobile domain wall.
The pattern of behavior is as follows.
A domain wall which is relatively flat and impurity-deficient moves
to the right uninhibited at first.
As it moves, however, it ``collects" impurities and eventually
becomes so overwhelmed that it practically stops moving.
Even so, the domain wall may become rough when fluctuations result
in an uneven impurity distribution \textit{along the wall} in the
vertical direction.
Portions of the wall which are impurity-rich move very little, while
impurity-deficient portions continue moving to the right.
As the resulting wall roughness increases, mobile,
impurity-deficient regions of the wall are often seen to ``engulf"
impurity-rich regions, leaving behind a pocket of impurities and a
relatively flat, impurity-deficient wall which travels to the right
virtually uninhibited (see Fig. \ref{rarEvent}).
The cycle continues.

Rationalizing hysteresis observations as a function of parameters
(such as the impurity-domain wall interaction strength $E_0$) was
the main goal of \cite{mikko}; indeed, a rationalization was
obtained by using the number of impurities on the domain wall, $N$,
as a coarse variable.
The selection of this variable was based on long observations of
system transient simulations and intuition.
Model reduction, in the form of an effective Fokker-Planck
description of the domain wall motion dynamics in terms of this
single coarse variable
\begin{eqnarray}
\frac{\partial P(N,t)}{\partial t} = -\frac{\partial}{\partial
N}\left[V(N)\, P(N,t) \right] + \\
\frac{\partial^2}{\partial N^2}\left[D(N) \, P(N,t) \right],
\nonumber
\end{eqnarray}
yielded an effective free energy surface with two ``wells": one
corresponding to domain wall motion, and one to stationarity.
$P(N,t)$ is here the probability that at time $t$, the number of
impurities on the domain wall is $N$; the quantities $V(N)$ and
$D(N)$ appearing in the effective Fokker-Planck equation are not
available in closed form, and were estimated from simulation data.

Changing parameters changed the relative depth and barrier height
between these wells, making hysteresis more or less apparent,
depending on the typical time scale of an observer.
This type of effective description was comparably successful over a
wide set of atomistic model parameter values \cite{mikko}.
This apparent success was, however, partially mitigated by the fact
that the one-dimensional reduced dynamic description was less
quantitative for small values of $N$ (see Fig.3 in \cite{mikko}).
This gradual deterioration of the predictive strength of the reduced
one-dimensional stochastic model strongly suggested the existence of
additional ``hidden variables", variables whose statistics do not
become quickly slaved to $N$ during simulation, in the low-$N$
regime.

In attempting to guess what such additional important coarse
observables might be, one might return to the observation of long
simulations, especially in the moderate wall impurity regime, when
the domain wall transitions from stationarity to motion (where, as
we discussed above, a type of roughening of the domain wall is often
observed).
%

\section{Data mining, diffusion maps and a distance metric \label{distmet}}

While physical understanding and intuition may suggest additional
physical observables for model reduction purposes (we already
mentioned observations of interface roughness, and we will return to
it below), a systematic, computer-assisted procedure for
detecting/suggesting such additional variables is clearly desirable.
The use of manifold-learning techniques (such as Isomap or Diffusion
Maps) has started to be explored in the context of reducing
atomistic models (see, e.g. \cite{cecilia,frewenchem,laing}).

We will employ the diffusion map approach here.
As described in the Appendix, we start with a reasonably extensive
ensemble of simulation snapshots, and with a scalar similarity
measure $W_{ij}$ of the ``closeness" between each pair ${i,j}$ of
these snapshots.
To achieve a low-dimensional embedding for a data set of $M$
individual $d$-dimensional vectors $\vec X_1$,...,$\vec X_M$ with
real entries, the Euclidean distance (possibly weighted) can help
construct such a similarity measure:
\begin{equation}
W_{ij} = \text{exp} \left[ - \left(\frac{||\vec X_i - \vec
X_j||}{\epsilon} \right)^2 \right].
\end{equation}
Here $\epsilon$ provides a separation scale beyond which the data
are considered effectively dissimilar.
An interpretation of this scale is that the Euclidean distance
between data points is ``meaningful" (meaningless) in determining
data similarity when it is smaller (larger) than $\epsilon$.
An ``inner" and an ``outer" distance play a role in the diffusion
map approach.
The inner distance (e.g. the Euclidean distance above) can be the
basis of a meaningful similarity measure between data points when it
is small; the diffusion map approach builds on such locally
meaningful similarity measures to construct a \textit{global}
distance, the so-called \textit{diffusion distance}, that
meaningfully quantifies data similarity even when the data are far
away from each other.
As a byproduct, the procedure also provides a new set of coordinates
(based on eigenfunctions of a diffusion operator, or, in discretized
form, on eigenvectors of a Markov matrix) which may prove useful in
embedding the data.
If we are lucky, a spectral gap will appear in the spectrum of the
diffusion operator, and a relatively small number of these new
coordinates can be used in reducing the dimensionality of the data
(and, thus, hopefully, of the model itself).
If the dimensionality of the data can indeed be reduced, the
diffusion distance is the Euclidean distance in the new space in
which the data is embedded.
The essence of the process is therefore to transform a
``trustworthy" local distance into a trustworthy global one in a new
set of coordinates; and while the Euclidean distance is an obvious
inner distance candidate, it is by no means the only one.

\begin{figure}
\includegraphics[height=102mm,width=130mm,keepaspectratio=false,angle=270]{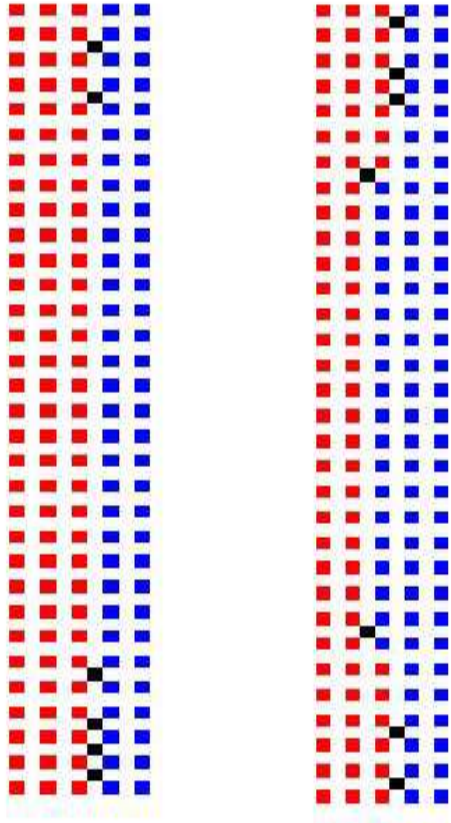}
\caption{Two snapshots exhibiting large pixel-by-pixel differences
which, however, are close dynamically: they are likely to transform
to each other via a small number of impurity diffusion steps and a
few quick, energetically favorable spin flips.\label{dynclose}}
\end{figure}

If physical understanding or intuition can suggest a locally
meaningful inner distance (beyond the obvious Euclidean norm), it
can serve as the basis of a successful reduction process.
Our data (our system snapshots) come in the form of 2048 ``pixels":
the 1024 spin lattice sites have either $+1$ or $-1$ entries, while
the 1024 interstitial sites have either $1$ (impurity present) or
$0$ entries.
At first thought, one might consider vectorizing each of the system
snapshots and using the Euclidean distance as an inner distance; yet
the entry values are clearly arbitrary, and one needs to devise some
relative weight for the spin and impurity entries.
Fig. \ref{dynclose} shows an example of two snapshots that have
large pixel-by-pixel differences, yet are close dynamically, in the
sense that a brief simulation can easily transform the one to the
other.
Clearly, coming up with a good inner distance for our data is
nontrivial.
We could possibly consider smoothing the lattice into spatial fields
(with the help of localized bump functions at each lattice and
interstitial site), or using Wasserstein-type (earthmover's
distance) metrics.
However, since we are interested in constructing a reduced
\textit{dynamic} model of the process, we decided to search for a
local distance that would incorporate some form of dynamic
similarity of snapshot pairs: how easy/likely it is for the direct
simulation to transform one to the other.

\begin{figure}
\includegraphics[height=170mm,width=130mm,keepaspectratio=false,angle=270]{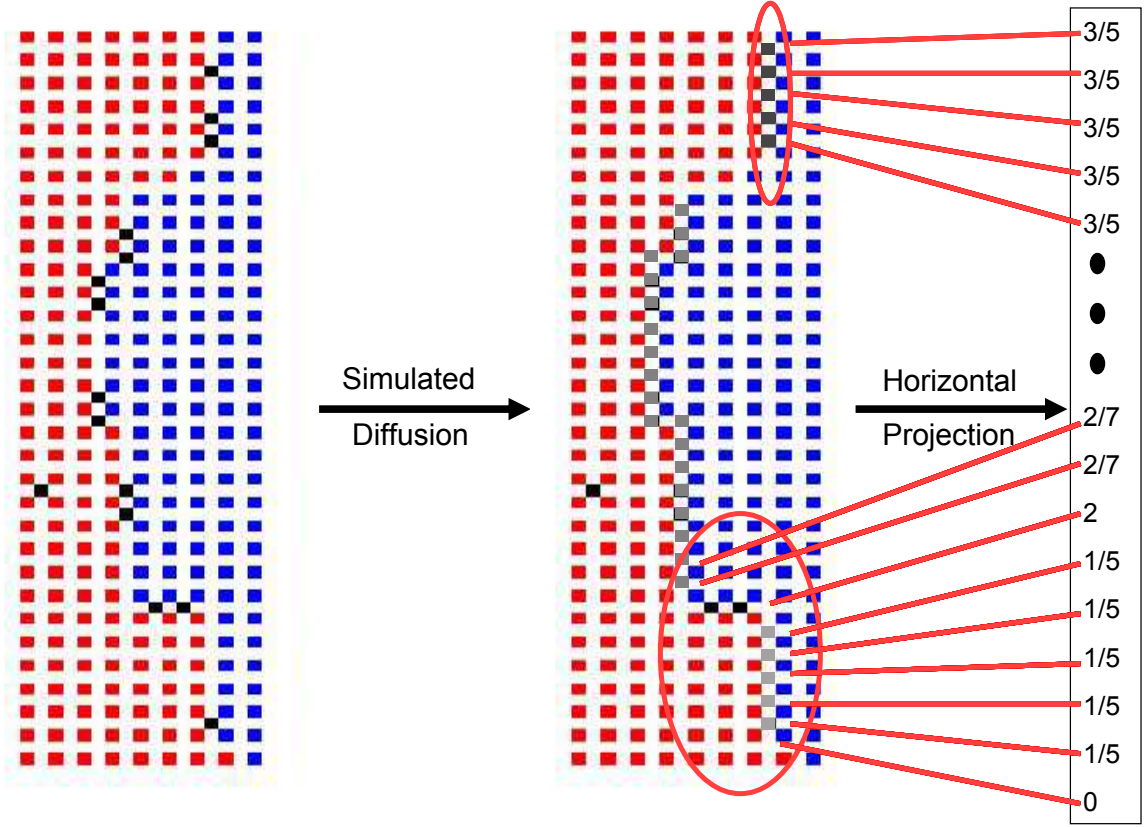}
\caption{The \textit{simulated diffusion} used in our similarity
measure construction. In a time interval that is short relative to
the domain wall motion timescale, impurities diffuse along the
boundary resulting in an apparent short-term equilibrium.
The only energetic barrier to this short-term motion is the presence
of domain wall corners (see text) .
After simulated diffusion, we add up the total amount of
\textit{diffused impurity} in each row and place the answer in the
corresponding entry of a vector (rightmost in the
figure).\label{simdiff}}
\end{figure}

We started by exploring the set of likely system snapshots through
``long" kinetic Monte Carlo simulations; we regularly sampled the
simulation, obtaining $M$ system snapshots, denoted $\{\mathbf{P}_i
\}^M_{i=1}$.
Inspection of the Hamiltonian shows that each impurity is free to
diffuse along horizontal or vertical straight segments of the domain
wall without energetic penalty and that these diffusion steps occur
over times very short relative to the domain wall evolution; hence,
two snapshots that differ only by such diffusion steps can be
considered dynamically close.
In a process we call \textit{simulated diffusion} (and only for the
purpose of ``off-line" computing our proposed inner distance) we
replace each impurity on a straight segment of the domain wall by a
``smeared out'' version of itself.
Specifically, we divide the domain wall into distinct straight
segments of varying lengths and determine the number of impurities
on each segment.
Then, on each segment, we distribute these impurities uniformly on
the interstitial sites (see Fig. \ref{simdiff}).
After this simulated diffusion, we ``summarize" our system by a
vector of length $32$, $\vec z$, adding up the fractions of
\textit{diffused impurity} in each row (again, see Fig.
\ref{simdiff}).
Entries of the vector with value greater than one correspond to
horizontal segments of the domain wall, and entries with values
between zero and one correspond to vertical segments; zero entries
may correspond to either.
%
We have found that even though this summary appears, at first sight,
to be an enormous loss of information about the system, it is
nevertheless a useful one: the behavior of the system over medium
time scales is not determined by the specific positions of domain
spins and impurities, but rather, by the overall shape of the domain
wall, i.e. its horizontal and vertical segments and how much
diffused impurity is present on each of these.
The vector contains exactly this type of information.

Two system snapshots $\mathbf{P}_1$ and $\mathbf{P}_2$ are (through
simulated diffusion) mapped to two $32$-long vectors $\vec z_1$ and
$\vec z_2$.
Because of the geometry (periodicity in the vertical ($y$)
direction, reflection symmetry around the $x$-``axis'') each vector
$\vec z_i$ is equivalent to (has zero distance from) 63 other
vectors obtained by index reflection (e.g. $[\vec z_i^{(1)}, \vec
z_i^{(2)} ,..., \vec z_i^{(31)}, \vec z_i^{(32)} ] \sim [\vec
z_i^{(32)}, \vec z_i^{(31)}, ..., \vec z_i^{(2)}, \vec z_i^{(1)}]$)
and cyclic index shifts (e.g. $[\vec z_i^{(1)}, \vec z_i^{(2)}, ...,
\vec z_i^{(31)}, \vec z_i^{(32)}] \sim [\vec z_i^{(k+1)},\vec
z_i^{(k+2)},...,\vec z_i^{(32)}, \vec z_i^{(1)},\vec
z_i^{(2)},...,\vec z_i^{(k)}]$); the set of all these equivalent
vectors is denoted as $\mathbf{Z}_i$.
The problem is also translationally invariant in the horizontal
direction; yet, our vector snapshot summary has automatically
accounted for this symmetry.
Our proposed inner distance $\mathbf{d}$ is then
\begin{equation}
\mathbf{d}(\mathbf{P}_1,\mathbf{P}_2) = \text{min}_{\vec z_2 \in
\mathbf{Z}_2} ||\vec z_1-\vec z_2|| = \text{min}_{\vec z_1 \in
\mathbf{Z}_1} ||\vec z_1-\vec z_2||,
\end{equation}
where for the rest of this paper, $||\,\,\, ||$ denotes the $L_2$
norm unless otherwise specified.
%
%
We will show below how we used this inner distance; we found it
quite robust in that small variations in its definition (e.g.
smearing impurities as a Gaussian, rather than uniformly along
straight segments) led to comparable diffusion map embeddings.


\section{The resulting two-dimensional description \label{resultdescr}}

\begin{figure}
\includegraphics[height=170mm,width=130mm,keepaspectratio=false,angle=270]{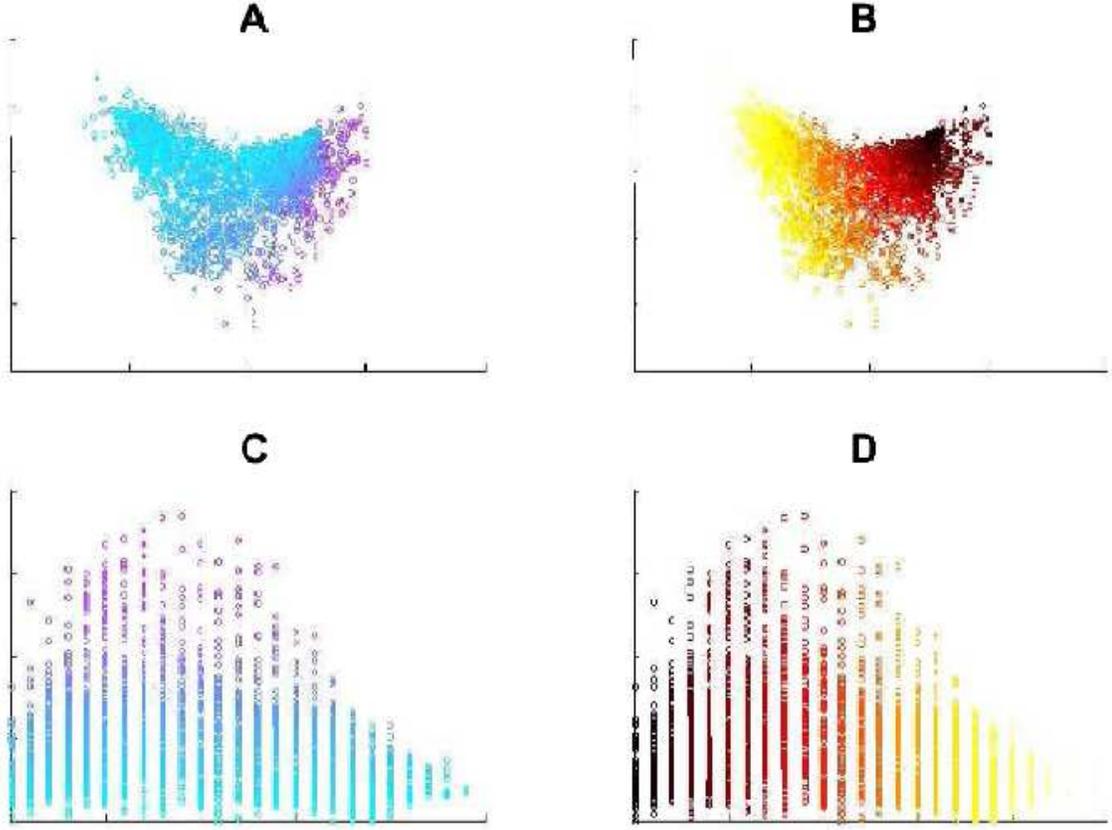}
\caption{Coarse variables: data mining and/or intuition.
Figures A and B show the data ensemble plotted in terms of their
components in the two leading (nontrivial) diffusion map
eigenvectors.
Figures C and D show the same data plotted now in terms of their
respective number of impurities on the domain wall $N$ (x-axis) and
their "roughness" $\mathbf{J}$ (y-axis, see text).
To assist in correlating the two representations, figures A and C
are colored by "roughness"; figures B and D are colored by number of
impurities on the domain wall.
These figures visually suggest that the determinant of the Jacobian
of the transformation from physical variables to diffusion map
coordinates is everywhere nonzero.\label{jac2d}}
\end{figure}

Using our inner distance, we computed (along the lines of the
Appendix) the eigenvalues of the Markov matrix resulting from the
ensemble of system snapshots.
In addition to the trivial eigenvalue at 1, two eigenvalues appear
significant, suggesting that we might be able to parameterize the
snapshots using their components in the corresponding two
eigenvectors, yielding a two-dimensional description.
The data plotted in terms of these two diffusion map coordinates are
shown in Figs. \ref{jac2d} A and \ref{jac2d} B.

\subsection{Relating diffusion map coordinates to physical observables}
Although the diffusion map approach is capable of discovering a
low-dimensional parameterization of our system snapshot data
ensemble which is mathematically optimal in a certain sense (e.g.
see \cite{coifman1} or \cite{belkin2003}), it does not provide
physical intuition about the embedding.
Thus, it makes sense to explore how the selected coordinates
correlate with physical observables.
Based on the intuition in \cite{mikko}, we test whether the
diffusion map variables correlate with our original single coarse
variable: the number of impurities on the domain wall.
Figure \ref{jac2d} D colors the data in terms of their domain wall
impurity number; a visual inspection of the ``color layering" in the
plot suggest a strong, but not perfect, correlation of this
observable with the first detected diffusion map variable.
Notice that the layering becomes increasingly slanted towards the
right, that is, towards low impurity counts.
This is consistent with \cite{mikko}: the number of impurities
provided there a good coarse variable for high impurity counts
(towards the left in the figure).
Yet the one-dimensional model deteriorated significantly at low
impurity counts (towards the right in the figure); this is precisely
what prompted our search for additional coarse variables.

Based on physical intuition, we expect a good second observable to
contain information associated not only with the number of domain
wall impurities, but also with the domain wall shape.
One way to quantify this might use the nonuniformities in the
impurity distribution along the domain wall length; we have
repeatedly observed that impurity-rich domain wall regions exhibit
high local curvature, forming apparent ``bubbles" that may be
engulfed, thus releasing the newly impurity-deficient wall.
We instead constructed a more direct measure of the wall roughness
associated with its high local curvature; we denote this measure as
$\mathbf{J}$ and compute it as follows.

Consider the sequence $\{\mathbf{F}_i\}_{i=1}^{31}$ of $32 \times
32$ spin lattices with perfectly vertical domain walls:
$\mathbf{F}_1$ has only one (the first, leftmost) column of $+1$
spins, and all its other columns are $-1$ spins; $\mathbf{F}_2$ has
two leftmost $+1$ spin columns, and so on.
For a system snapshot $\mathbf{P}$ with domain spin lattice entries
$\mathbf{S}$, $\mathbf{J}$ is found by minimizing (over the set of
lattices $\{\mathbf{F}_i\}_{i=1}^{31}$) the square of the Frobenius
norm
\begin{equation}
\mathbf{J}(\mathbf{S}) = \text{min}_{\{\mathbf{F}_i\}_{i=1}^{31}}
||\mathbf{F}_i-\mathbf{S}||_F^2.
\end{equation}

It might be, at first thought, surprising that the computation of
our roughness measure involves no impurity information whatsoever;
yet the placement of domain wall impurities and the overall domain
wall shape are highly correlated: a rough domain wall which contains
no impurities will very quickly (in the simulation) become flat.
The combination of the four panels (considering also their shading)
strongly suggests that there exists a bijection between the two
coordinates discovered through the diffusion map process and the two
coordinates arising from physical considerations.
Indeed, the Jacobian of the transformation between the two pairs of
coordinates does not become singular over the data ensemble, as we
have confirmed numerically.

\subsection{An effective dynamic model \label{mathEffDescr}}
Along the lines of \cite{mikko}, we attempt to describe the reduced
system dynamics in terms of an effective two-dimensional
Fokker-Planck equation,

\begin{eqnarray}
\frac{\partial P(\psi_1,\psi_2,t)}{\partial t} = - \sum_{i=1}^2
\frac{\partial}{\partial \psi_i}
\left[D_i^1(\psi_1,\psi_2)P(\psi_1,\psi_2,t) \right] + \\
\sum_{i,j=1}^2 \frac{\partial^2}{\partial \psi_i \partial \psi_j}
\left[D^2_{ij}(\psi_1,\psi_2)P(\psi_1,\psi_2,t) \right]. \nonumber
\end{eqnarray}

\begin{figure}
\includegraphics[height=170mm,width=80mm,keepaspectratio=false,angle=270]{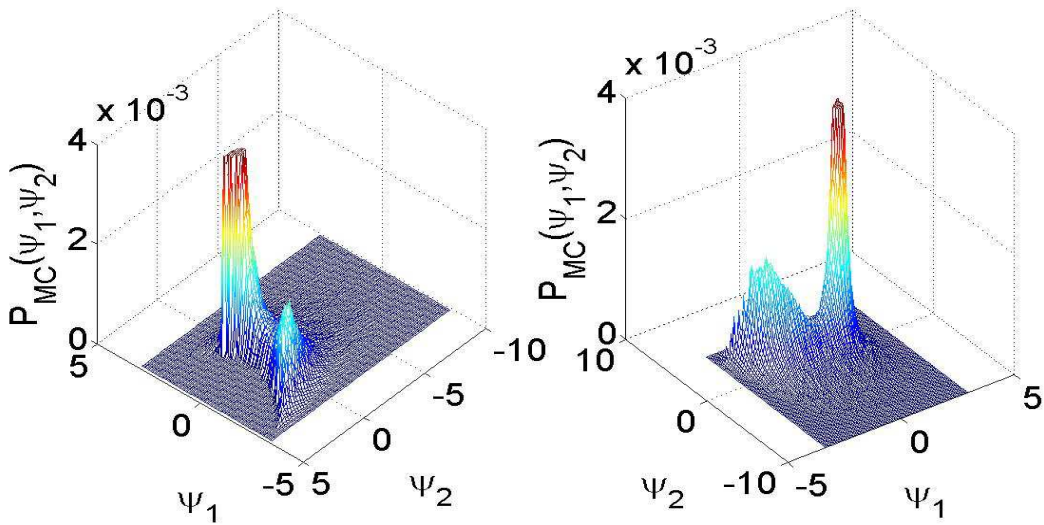}
\caption{Two views of the steady-state probability distribution
obtained by long Monte Carlo simulation viewed from two different
angles; the distribution is truncated at $P = 4 \times 10^{-3}$ but
extends approximately to the height $P = 8 \times 10^{-3}$.
\label{eqdistt}}
\end{figure}

Here the drifts and diffusion coefficients in the equation can be
estimated by $D_i^1(\psi_1,\psi_2) = \text{lim}_{\Delta t
\rightarrow 0} \mbox{$<\psi_i(t+\Delta t)-\psi_i(t)>/\Delta t$}$ and
$2 D_{ij}^2(\psi_1,\psi_2) = \text{lim}_{\Delta t \rightarrow 0}
\mbox{$<\left(\psi_i(t+\Delta t)-\psi_i(t)
\right)\left(\psi_j(t+\Delta t) - \psi_j \right)>/\Delta t$}$.
As in the one-dimensional case, setting $\frac{\partial
P(\psi_1,\psi_2,t)}{\partial t} = 0$ gives the steady state
probability distribution $P_{FP}(\psi_1,\psi_2)$.
If the effective Fokker-Planck model is accurate, $P_{FP}$ will be
well approximated by histograms constructed using long kinetic
Monte-Carlo simulation data, $P_{MC}(\psi_1,\psi_2)$; see Fig.
\ref{eqdistt}.

The $D_{ij}^2$ and $D_i^1$ may be estimated in several ways, two of
which are discussed here.
The first is to choose a particular point $(\psi_1,\psi_2)$ in
diffusion map space (actually, a small ``box" surrounding it),
locate several instances it arises in long kinetic Monte-Carlo
simulations, and then for each of these instances, record the
observed diffusion map coordinates $\Delta t$ later.
Averaging these results as shown above will provide a numerical
estimate of the $D_{ij}^2$ and $D_i^1$.

\begin{figure}
\includegraphics[height=170mm,width=130mm,keepaspectratio=false,angle=270]{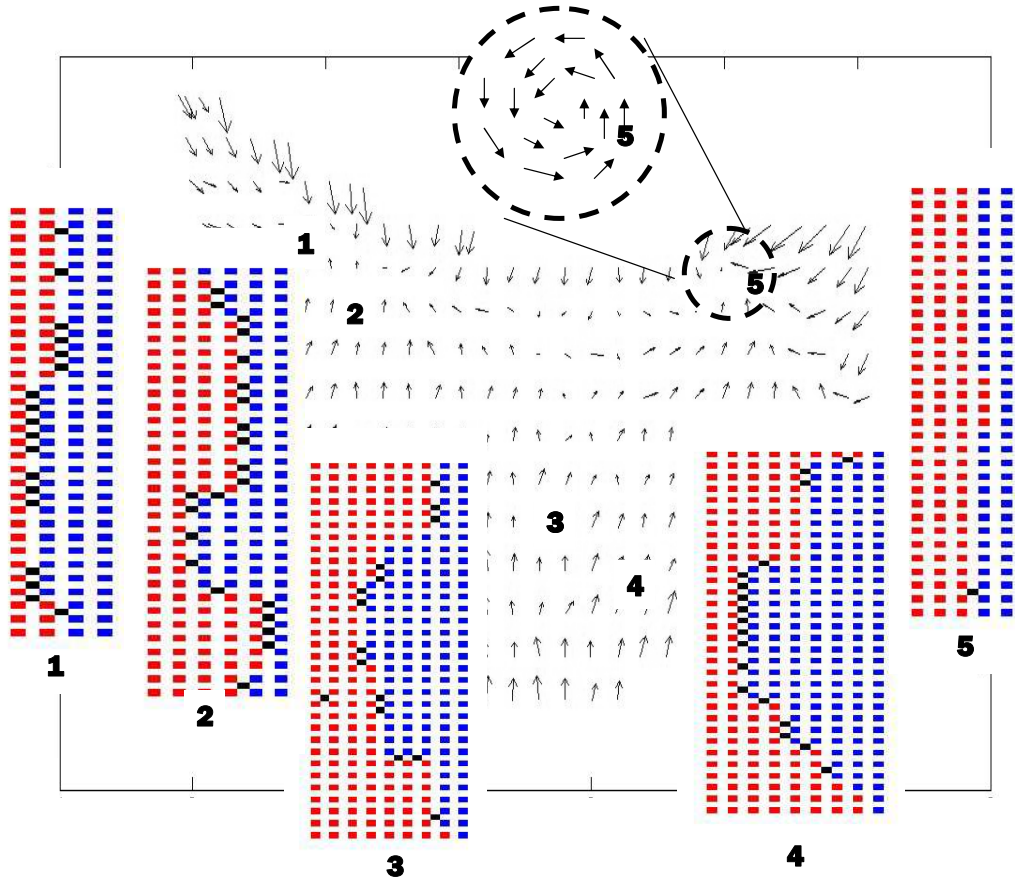}
\caption{The estimated effective drift vector field in diffusion map
coordinate space, along with representative system snapshots at five
selected locations to assist rationalizing the dynamics.
Snapshots 1 and 2 lie in a high-impurity, low-roughness region,
while 5 is located in a low-impurity, low-roughness region.
The domain wall of snapshot 3 will evolve towards the apparent
saddle point, while the wall in 4 appears to be increasing its
roughness, preparing to engulf its impurities. \label{flowmap}}
\end{figure}

\begin{figure}
\includegraphics[height=170mm,width=92mm,keepaspectratio=false,angle=270]{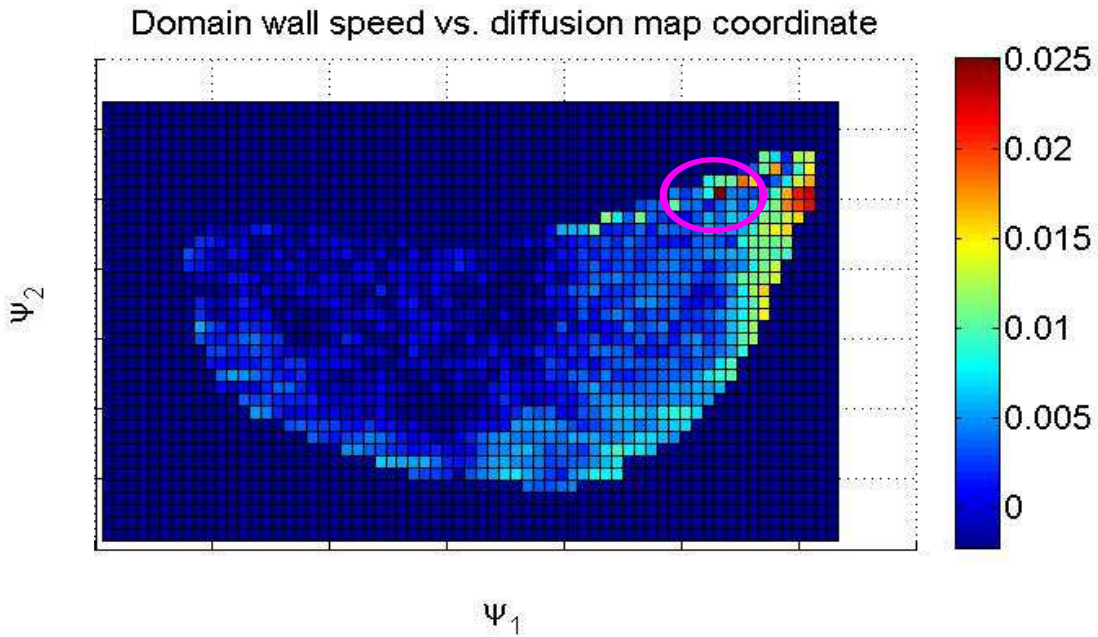}
\caption{Estimated domain wall speed as a function of the diffusion
map coordinates.
Velocity is in units of length per unit of time, where one unit of
length is defined as the width of a domain spin and the unit of time
was defined above.
High speeds occur when the domain wall is free of impurities (at the
center of the pink circle) and/or it has high roughness (taking Fig.
\ref{jac2d} into account).
%
%
%
\label{speedFig}}
\end{figure}

Fig. \ref{flowmap} shows the ``drift vector field", the plot of
$D_i^1(\psi_1,\psi_2)$ estimated this way.
At each point $(\psi_1,\psi_2)$ on the diffusion map plane, the
vector $\left[D_1^1(\psi_1,\psi_2),D_2^1(\psi_1,\psi_2)\right]$ is
plotted.
The estimation has been performed in diffusion map coordinate space;
we are, however, able to directly transform to the $\mathbf{J}$/$N$
space as illustrated in Fig. \ref{jac2d}; we have included in Fig.
\ref{flowmap} representative snapshots at selected points to assist
physical discussion of the drift dynamics.
Visual inspection of the drift vector field (ignoring, for the
moment, the stochastic component of the dynamics) shows exactly what
we might expect for this system: starting on the right (somewhere in
the circle, shown in the blowup) the domain wall remains relatively
flat as it gathers impurities; it eventually engulfs these
impurities by first becoming more rough.
There are two apparent steady states in this drift vector field: one
to the right, where an impurity-deficient domain wall ``slips" with
some apparent speed; and, one to the left, where an impurity-rich
domain wall ``sticks".
A saddle-type steady state can be seen for the drift vector field;
it lies in a region characterized by moderate roughness and moderate
number of wall impurities.
Here domain walls (again ignoring, for the moment, the stochastic
component of the dynamics) will, under small perturbations, either
go on to gather more impurities and ``stick", or they may become
more rough, engulf their impurities, and cycle back to low-impurity
fast-moving states.
Fig. \ref{speedFig} summarizes much of this information by plotting
estimated domain wall velocity vs. diffusion map coordinate.
$(\psi_1,\psi_2)$.

The Ising model variant we study here does not appear to be a
gradient system in two effective dimensions; observe, for instance,
the curl-intense region of the drift vector field in Fig.
\ref{flowmap}.
This is precisely the region which suggested the need for a second
coarse variable in \cite{mikko}.
This, unfortunately, does not allow us to simply and directly obtain
the steady-state probability distribution in terms of an effective
free energy that can be explicitly constructed from the estimated
Fokker-Planck coefficients.
We can, however, proceed to numerically solve, using standard
discretization techniques (e.g. finite elements), for the steady
state of the (numerically estimated) Fokker-Planck equation in
two-dimensional diffusion map coordinate space.

A second approach to estimating the Fokker-Planck equation
coefficients is motivated by the corresponding Langevin stochastic
differential equation and uses short bursts of kinetic Monte-Carlo
simulations appropriately initialized at particular values of the
diffusion map coordinates.
Indeed, if the diffusion map coordinates we constructed are
dynamically meaningful (which we certainly expect because our
similarity measure was dynamically motivated), initializing an
ensemble of short simulation bursts on demand at specified diffusion
map coordinates allows us to estimate the drift and diffusion
coefficients by processing the results of short burst simulations
through the above formulas.
A new computational challenge arises now: we may wish to initialize
the kinetic Monte Carlo simulations at points in diffusion map space
not included in our original data ensemble.
Furthermore, the simulations will produce new snapshots in physical
space which are also not included in our original data ensemble.
Clearly, for this approach to be viable, we need operators that
successfully translate information between physical and diffusion
map space.
These are the ``lifting" (coarse to fine, diffusion map to
spin/impurity lattice) and ``restriction" (fine to coarse,
spin/impurity lattice to diffusion map) operators of equation-free
computation \cite{man1,man2}.

\section{Lifting and Restriction \label{lifting}}

The \textit{restriction} operator, which finds the diffusion map
coordinates of a new system snapshot $\mathbf{P}_{\text{new}}$ that
was not part of the original data ensemble, is implemented via
Nystr\"om extension (see the Appendix).
It involves the computation of the similarity of the new data point
to the points of the original ensemble (possibly only the nearby
points).

\textit{Lifting}, the inverse problem of finding a new physical
initial condition with prescribed diffusion map coordinates, is
significantly more computationally involved than restriction;
because of the nonlinear nature of the diffusion-map based model
reduction, it is significantly more complicated than, say, reduction
using Principal Component Analysis.
Lifting is a ``one-to-many" operation; there are many physical
configurations that reduce to (practically) the same two diffusion
coordinate values.
Here the lifting problem is solved through a simulated annealing
algorithm \cite{sciencesimann}, which is modified by linking to the
kinetic Monte Carlo simulation as we now describe.
We note that each lift requires \textit{multiple} applications of
the restriction operator.

\subsection{The lifting procedure}

The only inputs to the lifting procedure are the two target
diffusion map coordinate values, $\psi_1^{\text{targ}}$ and
$\psi_2^{\text{targ}}$.
We start with an arbitrary trial system snapshot (possibly a
snapshot of the original ensemble whose diffusion map coordinates
are close to the target values) and evaluate its diffusion map
coordinates using Nystr\"om extension.
We attempt to modify this trial system snapshot by either changing
the number of wall impurities (preserving the total lattice impurity
count) or by flipping a small neighborhood, or ``block", of spins
close to the domain wall; we compute the modified diffusion map
coordinates again through Nystr\"om extension.
The modification is accepted if its diffusion map coordinates are
closer to the target values; if not, the change is accepted with
some probability that depends exponentially on the diffusion map
coordinate change.

\begin{figure}
\includegraphics[height=170mm,width=130mm,keepaspectratio=false,angle=270]{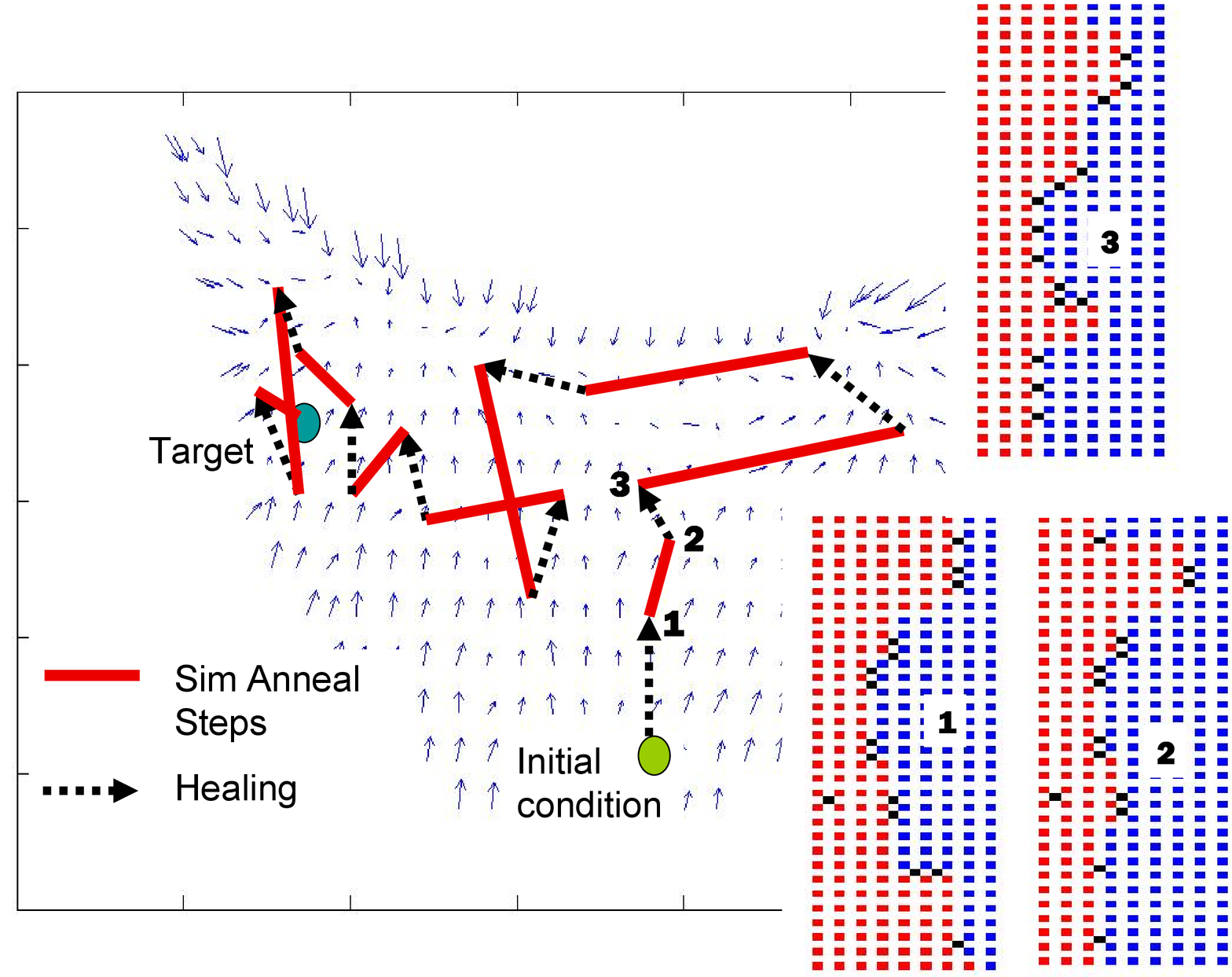}
\caption{Lifting using simulated annealing alternating with
simulation-induced healing.
Only accepted simulated annealing steps are shown.
The perturbation scale (flipped spin block size), annealing
temperature, and healing time become increasingly smaller as the
target is approached.
Brief healing paths tend to flow along drift vector field
trajectories. \label{figsimann}}
\end{figure}

The concrete steps are as follows (see Fig. \ref{figsimann}):
\begin{enumerate}
\item Start with a random initial system snapshot.
\item Evolve (through kinetic Monte Carlo) the system briefly, allowing it
to \textit{heal} for a relatively short amount of simulation time.
Here, we choose 20 units of time.
Denote the healed system snapshot as $\mathbf{P}_{\text{curr}}$.
Healing is intended to get rid of improbable snapshot features
artificially generated by the simulated annealing steps.
Determine the diffusion map coordinates of this healed snapshot
using Nystr\"om extension.
In order to decide about accepting the modification or not, we need
to prescribe an \textit{annealing temperature}, which we choose here
to be $T = ||\Psi_2(\mathbf{P}_{\text{curr}})-[\psi_1^{\text{targ}},
\psi_2^{\text{targ}}]||$ (see Appendix for meaning of $\Psi_2$).
Depending on how far away from the desired coordinate values we are
(i.e., depending on the annealing temperature) we allow for bigger
perturbations; we set the height of the block of spins to be flipped
to  $H = \text{min}\left(18,\text{max}(3,\text{ceil}(18\,
U_1([0,1])\, \sqrt T))\right)$, and its width to $W =
\text{min}\left(6,\text{max}(1,\text{ceil}(6\, U_2([0,1])\, \sqrt
T))\right)$, where $U_1$ and $U_2$ are uniform random variables.
This way, for moderate $T$, $HW \approx 27T$.
$H$ is chosen longer than $W$ on average, in order to keep the
perturbed domain wall flatter in the steps that follow.
These choices are motivated by observing statistics of fluctuations
during the system dynamic evolution; the procedure is quite robust
to the details of such choices.
\item \label{trialsnap} Generate a new trial snapshot $\mathbf{P}_{\text{trial}}$ by either
\begin{enumerate}
\item  Adding an impurity to an empty spot on the domain
wall boundary,  or by removing an impurity on the domain wall
boundary (keeping the total impurity count constant).
\item Replacing a block of $+1$ spins intersecting the domain wall with $-1$ spins.
The block is $H$ rows high and $W$ rows wide, and its placement is
selected randomly and uniformly from the set of all possible
placements that intersect the domain wall.
Without intersection with the domain wall, flipping the spin block
would have no effect; blocks taken in regions entirely to the right
are already comprised of $-1$ spins, while blocks taken in regions
entirely to the left of the domain wall would flip back almost
immediately.
If the selected block contains impurities, these impurities are
projected to the left, so that they remain on the new domain wall.
The system is allowed to heal for $HW$ units of time.
\end{enumerate}
\item The new trial snapshot is accepted
(we set $\mathbf{P}_{\text{curr}} = \mathbf{P}_{\text{trial}}$) if
its diffusion map coordinates (computed through Nystr\"om extension)
are closer to the target diffusion map coordinates, or with
probability
\begin{eqnarray}
P = \text{exp} [-\frac{1}{T}(||\Psi_2(\mathbf{P}_{\text{trial}})-[\psi_1^{\text{targ}},\psi_2^{\text{targ}}]||-\\
||\Psi_2(\mathbf{P}_{\text{curr}})-[\psi_1^{\text{targ}},\psi_2^{\text{targ}}]||)]
\nonumber
\end{eqnarray}
otherwise.
\item The current annealing temperature is set to $T =
||\Psi_2(\mathbf{P}_{\text{curr}})-[\psi_1^{\text{targ}},\psi_2^{\text{targ}}]||$,
and $H$ and $W$ are reset accordingly.
If $T<\rho$, exit; $\mathbf{P}_{\text{curr}}$ is the desired
snapshot; otherwise, go back to step \ref{trialsnap}.
In our experience $\rho = 0.05$ gives satisfactory snapshots, close
enough to the target diffusion map variables.
\end{enumerate}

The healing steps, interspersed with the simulated annealing steps,
appear to be important in giving plausible physical snapshots with
the desired diffusion map coordinates (i.e. snapshots whose
diffusion map coordinates evolve typically slowly upon further
simulation).
There are many physical snapshots which (through Nystr\"om
extension) possess the same two diffusion map coordinates (we have
already said that lifting is a one-to-many operation).
What our simulations suggest is that, among these snapshots, we can
find both ``mature" ones, for which further diffusion map coordinate
evolution is slow, and ``unhealed" ones, for which further diffusion
map coordinate evolution quickly brings us to a different ``mature"
point on our two-dimensional manifold with different diffusion map
coordinates.
Our lifting process typically takes about 30 cycles, each requiring
a Nystr\"om extension.

There are a number of more or less \textit{ad hoc} choices in the
above algorithm, such as the type and scale of perturbations chosen,
the selection of the annealing temperature, the size of $H$ and $W$,
the healing time etc.
Having to make such choices is part of any simulated annealing
algorithm; we have found that our results are quite robust to
modifications in the specific choices presented above.

\subsection{Testing the lifting operator}
The variables detected by the diffusion map process are useful in
providing a good parameterization of the snapshot data; plotting the
data in terms of these coordinates can provide very helpful visual
summaries of the overall dynamics (e.g. Fig. \ref{flowmap}), and can
even help develop intuition and useful hypotheses about the system.
If a good lifting operator can be constructed, however, these
coordinates can be even more useful: they can assist in the design
of computational experiments that extract system-level information
from the problem.
One might wait, for example, for a long kinetic Monte Carlo
simulation to sample enough domain wall pinning/depinning events to
estimate the corresponding characteristic time; alternatively, short
bursts of kinetic Monte Carlo simulations can be designed, by
appropriately initializing them through lifting at selected
diffusion map coordinates, to extract this information exploiting
the effective Langevin (or Fokker-Planck) model discussed above.

Our variables are dynamically meaningful if a Langevin (or
associated Fokker-Planck) dynamic model in terms of only these
variables is successful in describing the problem (if, for example,
the Fokker-Planck coefficients estimated through long kinetic Monte
Carlo simulations and the ones estimated through short bursts
following a lifting procedure are practically the same).
We expect this to be the case since our similarity measure was
dynamically motivated.
Tables  \ref{liftingTableDrift} and \ref{liftingTableDiff} show
precisely such a test of our two diffusion map variables along with
the correct performance of our lifting operator: Fokker-Planck
coefficient estimates using each of the two approaches at seven
selected locations are favorably compared.
\begin{table}
\caption{Drift coefficients computed for certain target diffusion
map coordinates via the two methods discussed in the text.  Only one
of the two drift coefficients is shown. ``MC'' stands for ``Monte
Carlo'', while ``SB'' stands for ``short
burst''.\label{liftingTableDrift}}
\begin{ruledtabular}
\begin{tabular}{c c c c}
$\psi_1^{\text{targ}}$ & $\psi_2^{\text{targ}}$ & $D_{1\,MC}^1$ & $D_{1\,SB}^1$ \\
\hline
-1.65 & -1.00 & -1.76e-2 $\pm$ 4.3e-4 & -1.80e-2 $\pm$ 2.1e-4\\
1.25 & 1.00 & -7.81e-3 $\pm$ 3.2e-4 & -7.63e-2 $\pm$ 1.6e-4\\
-0.20 & -0.60 & -4.43e-3 $\pm$ 5.1e-4 & -4.20e-3 $\pm$ 2.5e-4\\
0.75 & 1.60 & 5.20e-2 $\pm$ 6.9e-4 & 5.29e-2 $\pm$ 3.4e-4\\
1.00 & 0.20 & -8.75e-3 $\pm$ 2.7e-4 & -8.81e-3 $\pm$ 1.3e-4\\
0.90 & -0.40 & -8.52e-3 $\pm$ 5.4e-5 & -8.49e-3 $\pm$ 3.2e-5\\
1.50 & -0.60 & -1.49e-1 $\pm$ 1.6e-3 & -1.51e-1 $\pm$ 7.76e-4\\
\end{tabular}
\end{ruledtabular}
\end{table}
\begin{table}
\caption{Diffusion coefficients computed for certain target
diffusion map coordinates via the two methods explained above. To
save space, only one of the three diffusion coefficients is shown.
Again, ``MC'' stands for ``Monte Carlo'', while ``SB'' stands for
``short burst''.\label{liftingTableDiff}}
\begin{ruledtabular}
\begin{tabular}{c c c c}
$\psi_1^{\text{targ}}$ & $\psi_2^{\text{targ}}$ & $D^2_{11\,MC}$ & $D^2_{11\,SB}$ \\
\hline
-1.65 & -1.00 & 4.52e-4 $\pm$ 6.2e-6 & 4.39e-4 $\pm$ 2.8e-6\\
1.25 & 1.00 & 2.53e-4 $\pm$ 3.3e-6 & 2.46e-4 $\pm$ 1.4e-6\\
-0.20 & -0.60 & 6.48e-4 $\pm$ 8.5e-6 & 6.29e-4 $\pm$ 4.3e-6\\
0.75 & 1.60 & 1.20e-3 $\pm$ 1.6e-5 & 1.14e-3 $\pm$ 5.7e-6\\
1.00 & 0.20 & 1.84e-4 $\pm$ 2.2e-6 & 1.78e-4 $\pm$ 1.2e-6\\
0.90 & -0.40 & 7.25e-6 $\pm$ 9.5e-7 & 1.03e-5 $\pm$ 7.2e-7\\
1.50 & -0.60 & 6.20e-3 $\pm$ 8.1e-5 & 6.02e-3 $\pm$ 3.8e-5\\
\end{tabular}
\end{ruledtabular}
\end{table}
The agreement is good, but not perfect:
possible reasons for the discrepancies range from the need to
collect longer data and more short replica simulations, to the
details of the estimation algorithm and even the possible inadequacy
of a white noise model in the effective Langevin description.
Tightening the $\rho = 0.05$ requirement in the simulated annealing,
or keeping more than two variables might improve the agreement.
Systematically testing the adequacy of the effective Fokker-Planck
description and improving the features of the above approach is a
vast mathematical and computational task that we do not attempt
here; our main point has been to suggest the use of diffusion map
coordinates as the variables of choice in this reduced modeling
context.

\section{Summary and Conclusions}

In this work we demonstrated the use of nonlinear manifold learning
techniques, and, in particular, diffusion maps, to obtain an
effective description for a complex, high-dimensional, stochastic
dynamical system.
Use of physical intuition and symmetries helped us formulate a
notion of a good local similarity measure between detailed system
snapshots.
Using a diffusion map, we obtained and validated a two-dimensional
effective description, and then we implemented a lifting procedure
that allowed us to obtain detailed physical system states consistent
with prescribed diffusion map coordinate values.
We also searched for and proposed two physical variables whose
relation to the corresponding diffusion map coordinates is a
bijection.
This helped the development of intuition about the important
features of the system.
Finally, we discussed the qualitative behavior of the system
exploiting the new low-dimensional description.

The computation of the diffusion map coordinates was performed
``off-line", before this entire process began.
The long-term goal is to use diffusion maps adaptively in an on-line
setting.
The idea is that a more local diffusion map parameterization, based
on local system observations, can be used to guide (i.e. bias)
further system simulation and collection of information by
suggesting ``interesting" new coarse initial conditions.
Such new simulations would then be used to continually modify and
extend local diffusion map coordinates into unexplored regions of
phase space.
For such a program to be implemented, a robust, automated and
computationally efficient lifting algorithm is clearly necessary.

\section{Acknowledgements}
This work was partially supported by the Department of Energy
Computational Science Graduate Fellowship (BES) and the US DOE
(IGK).

\appendix*
  \begin{center}
    {\bf APPENDIX}
  \end{center}

To construct a low-dimensional embedding for a data set of $M$
individual $d$-dimensional real vectors, $\vec X_1$,...,$\vec X_M$,
we start with a similarity measure between each pair of vectors
$\vec X_i$, $\vec X_j$.  The similarity measure is a nonnegative
quantity $W_{ij} = W_{ji}$ satisfying certain additional
``admissibility conditions" \cite{coifman1}.
As a concrete example, consider using a Gaussian similarity measure
(based on the standard $L_2$ norm):
\begin{equation}
W_{ij} = \text{exp} \left[ - \left(\frac{||\vec X_i - \vec
X_j||}{\epsilon} \right)^2 \right].
\end{equation}
A weighted Euclidean norm may be chosen over the standard $L_2$ norm
in situations where the values of different components of $\vec X$
may vary over disparate orders of magnitude.
$\epsilon$ defines a characteristic scale which quantifies the
``locality'' of the neighborhood within which Euclidean distance can
be used as the basis of a meaningful similarity measure
\cite{coifman1}. A systematic approach to determining appropriate
$\epsilon$ values is discussed below.
Next, we define the diagonal matrix $\mathbf{D}$ by
\begin{equation}
D_{ii} = \sum^M_{k=1} W_{ik},
\end{equation}
and then we compute the first few right eigenvectors and eigenvalues
of the stochastic matrix
\begin{equation}
\mathbf{K} = \mathbf{D}^{-1} \mathbf{W}.
\end{equation}
In MATLAB, for instance, this can be done with the command
$[\mathbf{V},\mathbf{D}] = \mbox{eigs}(\mathbf{K},N_{evl})$, where
$N_{evl}$ is the number of top eigenvalues we wish to keep (we
typically are only interested in the first few).
It is important to note that it is not required that our data be in
vector format; in the end, all we need to apply the diffusion map
approach is a pairwise similarity measure $W_{ij}$.

This gives a set of real eigenvalues $\lambda_0 \geq \lambda_1 \geq
... \geq \lambda_{M-1} \geq 0$ with corresponding eigenvectors
$\{\vec \psi_j\}_{j=0}^{M-1}$.
Since $\mathbf{K}$ is stochastic, $\lambda_0 = 1$ and $\vec \psi_0 =
[1\, 1 \, ... \, 1]^T$.
The $n$-dimensional representation of a particular $d$-dimensional
data vector, $\vec X_i$, is given by the \textit{diffusion map}
$\Psi_n:\, \vec X \longrightarrow \mathbf{R}^n$, where
\begin{equation}
\Psi_n(\vec X_i) = [\vec \psi_1^{(i)}, \vec \psi_2^{(i)}, ..., \vec
\psi_n^{(i)}],
\end{equation}
a mapping which is only defined on the $M$ recorded data vectors (in
order for Euclidean distance in diffusion map space to actually
equal diffusion distance, mentioned above, the diffusion map must be
instead scaled as $\Psi_n(\vec X_i) = [\lambda_1 \vec \psi_1^{(i)},
\lambda_2 \vec \psi_2^{(i)}, ..., \lambda_n \vec \psi_n^{(i)}]$).
In other words, the vector $\vec X_i$ is mapped to the vector whose
first component is the $i$th component of the first nontrivial
eigenvector, whose second component is the $i$th component of the
second nontrivial eigenvector, etc.
If, for instance, a gap in the eigenvalue spectrum is observed
between eigenvalues $\lambda_t$ and $\lambda_{t+1}$, then $\Psi_t$
gives a good low-dimensional representation of the data
set\cite{belkin2003,coiffp}.
It is interesting that if the data come from a Markovian stochastic
process, the eigenvectors and eigenvalues are approximations to the
eigenfunctions and eigenvalues of the corresponding backward
Fokker-Planck operator \cite{coiffp}.

\begin{figure}
\includegraphics[height=170mm,width=130mm,keepaspectratio=false,angle=270]{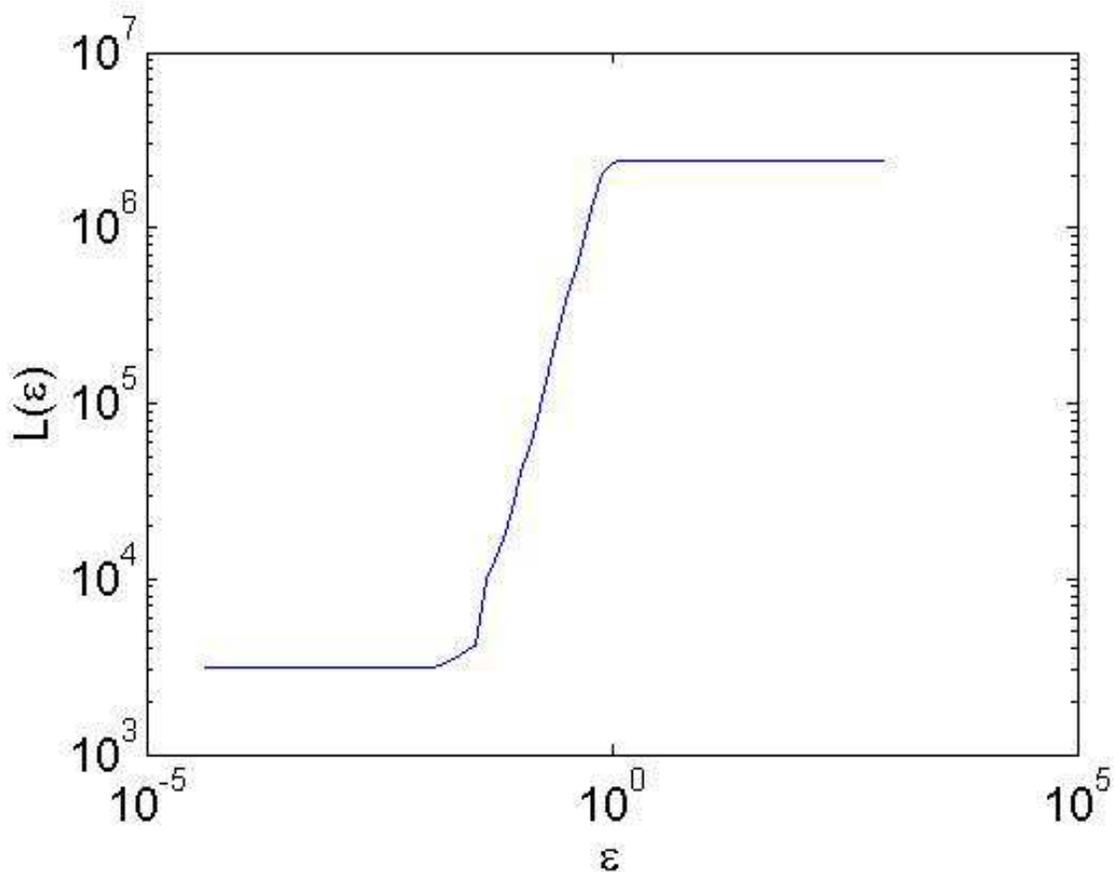}
    \caption{For $M$ data vectors, $\lim_{\epsilon \to \infty} L(\epsilon)$ is
    simply $M^2$ and $\lim_{\epsilon \to 0} L(\epsilon)$ is
    simply $M$.  These two asymptotes are connected, however, by a sloped, approximately
    straight line.  $\epsilon$ values chosen from this regime are appropriate.}
    \label{fig:loglog}
\end{figure}

We choose the value of $\epsilon$ used in the diffusion map
computation by invoking the \textit{correlation dimension} utilized
in \cite{loglog}.
The assumption here is that the volume of an $n$-dimensional set
scales with any characteristic length $s$ as $s^n$; for relatively
uniform sampling one might expect the number of neighbors less than
$s$ apart to scale similarly.
We first compute all pairwise distances.
Fig. \ref{fig:loglog} shows the total number of pairwise distances
less than $\epsilon$; it is clear that an asymptote will arise at
large $\epsilon$ ($M^2$, where $M$ is the number of points) and at
small $\epsilon$ ($M$).
In the figure, the two asymptotes are smoothly connected by an
approximately straight line;  the slope of this line suggests the
correct dimensionality for our data set (here, two).
The range of $\epsilon$ values corresponding to this straight
segment are all acceptable in our diffusion map computations: here
any value of $\epsilon$ between approximately $10^{-\frac{3}{2}}$
and $10^{-\frac{1}{2}}$ may be used.

The problem of finding the diffusion map coordinates of a
\textit{new} $d$-dimensional vector is solved with the Nystr\"om
extension.
The first step is to compute all distances
$\{\mathbf{d}_{i\,\text{new}}\}_{i=1}^M$ between our new vector and
the $M$ vectors in our data set, and set $W_{i\, \text{new}} =
\text{exp} \left[ - \left(
\frac{\mathbf{d}_{i\,{\text{new}}}}{\epsilon} \right)^2 \right]$;
here we used the standard Euclidean norm for $\mathbf{d}$.
Setting
\begin{equation}
K_{i \, \text{new}} = \left(\sum_{k=1}^M W_{k\,
\text{new}}\right)^{-1} W_{i\, \text{new}},
\end{equation}
the $j\text{th}$ diffusion map coordinate of the new vector
$\mathbf{P}_{\text{new}}$ is given by
\begin{equation}
\vec \psi_j^{\text{new}} = \frac{1}{\lambda_j} \sum_{i=1}^M K_{i \,
\text{new}} \vec \psi_j^{(i)}.
\end{equation}


\bibliography{Paper1008}

\begin{thebibliography}{19}
\expandafter\ifx\csname natexlab\endcsname\relax\def\natexlab#1{#1}\fi
\expandafter\ifx\csname bibnamefont\endcsname\relax
  \def\bibnamefont#1{#1}\fi
\expandafter\ifx\csname bibfnamefont\endcsname\relax
  \def\bibfnamefont#1{#1}\fi
\expandafter\ifx\csname citenamefont\endcsname\relax
  \def\citenamefont#1{#1}\fi
\expandafter\ifx\csname url\endcsname\relax
  \def\url#1{\texttt{#1}}\fi
\expandafter\ifx\csname urlprefix\endcsname\relax\def\urlprefix{URL }\fi
\providecommand{\bibinfo}[2]{#2}
\providecommand{\eprint}[2][]{\url{#2}}

\bibitem[{\citenamefont{Haataja et~al.}(2004)\citenamefont{Haataja, Srolovitz,
  and Kevrekidis}}]{mikko}
\bibinfo{author}{\bibfnamefont{M.}~\bibnamefont{Haataja}},
  \bibinfo{author}{\bibfnamefont{D.}~\bibnamefont{Srolovitz}},
  \bibnamefont{and} \bibinfo{author}{\bibfnamefont{I.~G.}
  \bibnamefont{Kevrekidis}}, \bibinfo{journal}{Phys. Rev. Lett.}
  \textbf{\bibinfo{volume}{92}}, \bibinfo{pages}{160603}
  (\bibinfo{year}{2004}).

\bibitem[{\citenamefont{Mendelev et~al.}(2001)\citenamefont{Mendelev,
  Srolovitz, and E}}]{mend2001}
\bibinfo{author}{\bibfnamefont{M.~I.} \bibnamefont{Mendelev}},
  \bibinfo{author}{\bibfnamefont{D.~J.} \bibnamefont{Srolovitz}},
  \bibnamefont{and} \bibinfo{author}{\bibfnamefont{W.}~\bibnamefont{E}},
  \bibinfo{journal}{Philos. Mag. A} \textbf{\bibinfo{volume}{81}},
  \bibinfo{pages}{2243} (\bibinfo{year}{2001}).

\bibitem[{\citenamefont{Cahn}(1962)}]{cahn62}
\bibinfo{author}{\bibfnamefont{J.~W.} \bibnamefont{Cahn}},
  \bibinfo{journal}{Acta Metall.} \textbf{\bibinfo{volume}{10}},
  \bibinfo{pages}{789} (\bibinfo{year}{1962}).

\bibitem[{\citenamefont{Tybell et~al.}(2002)\citenamefont{Tybell, Paruch,
  Giamarchi, and Triscone}}]{tybell2002}
\bibinfo{author}{\bibfnamefont{T.}~\bibnamefont{Tybell}},
  \bibinfo{author}{\bibfnamefont{P.}~\bibnamefont{Paruch}},
  \bibinfo{author}{\bibfnamefont{T.}~\bibnamefont{Giamarchi}},
  \bibnamefont{and} \bibinfo{author}{\bibfnamefont{J.-M.}
  \bibnamefont{Triscone}}, \bibinfo{journal}{Phys. Rev. Lett.}
  \textbf{\bibinfo{volume}{89}}, \bibinfo{pages}{097601}
  (\bibinfo{year}{2002}).

\bibitem[{\citenamefont{H\"ahner et~al.}(2002)\citenamefont{H\"ahner,
  Ziegenbein, Rizzi, and Neuh\"auser}}]{hahner}
\bibinfo{author}{\bibfnamefont{P.}~\bibnamefont{H\"ahner}},
  \bibinfo{author}{\bibfnamefont{A.}~\bibnamefont{Ziegenbein}},
  \bibinfo{author}{\bibfnamefont{E.}~\bibnamefont{Rizzi}}, \bibnamefont{and}
  \bibinfo{author}{\bibfnamefont{H.}~\bibnamefont{Neuh\"auser}},
  \bibinfo{journal}{Phys. Rev. B} \textbf{\bibinfo{volume}{65}},
  \bibinfo{pages}{134109} (\bibinfo{year}{2002}).

\bibitem[{\citenamefont{Cule and Hwa}(1996)}]{cule96}
\bibinfo{author}{\bibfnamefont{D.}~\bibnamefont{Cule}} \bibnamefont{and}
  \bibinfo{author}{\bibfnamefont{T.}~\bibnamefont{Hwa}},
  \bibinfo{journal}{Phys. Rev. Lett.} \textbf{\bibinfo{volume}{77}},
  \bibinfo{pages}{278} (\bibinfo{year}{1996}).

\bibitem[{\citenamefont{Belkin and Niyogi}(2003)}]{belkin2003}
\bibinfo{author}{\bibfnamefont{M.}~\bibnamefont{Belkin}} \bibnamefont{and}
  \bibinfo{author}{\bibfnamefont{P.}~\bibnamefont{Niyogi}},
  \bibinfo{journal}{Neural Comput.} \textbf{\bibinfo{volume}{15}},
  \bibinfo{pages}{1373} (\bibinfo{year}{2003}).

\bibitem[{\citenamefont{Coifman
  et~al.}(2005{\natexlab{a}})\citenamefont{Coifman, Lafon, Lee, Maggioni,
  Nadler, Warner, and Zucker}}]{coifman1}
\bibinfo{author}{\bibfnamefont{R.}~\bibnamefont{Coifman}},
  \bibinfo{author}{\bibfnamefont{S.}~\bibnamefont{Lafon}},
  \bibinfo{author}{\bibfnamefont{A.}~\bibnamefont{Lee}},
  \bibinfo{author}{\bibfnamefont{M.}~\bibnamefont{Maggioni}},
  \bibinfo{author}{\bibfnamefont{B.}~\bibnamefont{Nadler}},
  \bibinfo{author}{\bibfnamefont{F.}~\bibnamefont{Warner}}, \bibnamefont{and}
  \bibinfo{author}{\bibfnamefont{S.}~\bibnamefont{Zucker}},
  \bibinfo{journal}{PNAS} \textbf{\bibinfo{volume}{102}}, \bibinfo{pages}{7426}
  (\bibinfo{year}{2005}{\natexlab{a}}).

\bibitem[{\citenamefont{Coifman
  et~al.}(2005{\natexlab{b}})\citenamefont{Coifman, Lafon, Lee, Maggioni,
  Nadler, Warner, and Zucker}}]{coifman2}
\bibinfo{author}{\bibfnamefont{R.}~\bibnamefont{Coifman}},
  \bibinfo{author}{\bibfnamefont{S.}~\bibnamefont{Lafon}},
  \bibinfo{author}{\bibfnamefont{A.}~\bibnamefont{Lee}},
  \bibinfo{author}{\bibfnamefont{M.}~\bibnamefont{Maggioni}},
  \bibinfo{author}{\bibfnamefont{B.}~\bibnamefont{Nadler}},
  \bibinfo{author}{\bibfnamefont{F.}~\bibnamefont{Warner}}, \bibnamefont{and}
  \bibinfo{author}{\bibfnamefont{S.}~\bibnamefont{Zucker}},
  \bibinfo{journal}{PNAS} \textbf{\bibinfo{volume}{102}}, \bibinfo{pages}{7432}
  (\bibinfo{year}{2005}{\natexlab{b}}).

\bibitem[{\citenamefont{Tenenbaum et~al.}(2000)\citenamefont{Tenenbaum,
  de~Silva, and Langford}}]{tenen2000}
\bibinfo{author}{\bibfnamefont{J.}~\bibnamefont{Tenenbaum}},
  \bibinfo{author}{\bibfnamefont{V.}~\bibnamefont{de~Silva}}, \bibnamefont{and}
  \bibinfo{author}{\bibfnamefont{J.}~\bibnamefont{Langford}},
  \bibinfo{journal}{Science} \textbf{\bibinfo{volume}{290}},
  \bibinfo{pages}{2319} (\bibinfo{year}{2000}).

\bibitem[{\citenamefont{Lafon and Lee}(2006)}]{lafonlee}
\bibinfo{author}{\bibfnamefont{S.}~\bibnamefont{Lafon}} \bibnamefont{and}
  \bibinfo{author}{\bibfnamefont{A.~B.} \bibnamefont{Lee}},
  \bibinfo{journal}{Pattern Analysis and Machine Intelligence, IEEE
  Transactions on} \textbf{\bibinfo{volume}{28}}, \bibinfo{pages}{1393}
  (\bibinfo{year}{2006}),
  \urlprefix\url{http://dx.doi.org/10.1109/TPAMI.2006.184}.

\bibitem[{\citenamefont{Das et~al.}(2006)\citenamefont{Das, Moll, Stamati,
  Kavraki, and Clementi}}]{cecilia}
\bibinfo{author}{\bibfnamefont{P.}~\bibnamefont{Das}},
  \bibinfo{author}{\bibfnamefont{M.}~\bibnamefont{Moll}},
  \bibinfo{author}{\bibfnamefont{H.}~\bibnamefont{Stamati}},
  \bibinfo{author}{\bibfnamefont{L.}~\bibnamefont{Kavraki}}, \bibnamefont{and}
  \bibinfo{author}{\bibfnamefont{C.}~\bibnamefont{Clementi}},
  \bibinfo{journal}{PNAS} \textbf{\bibinfo{volume}{103}}, \bibinfo{pages}{9885}
  (\bibinfo{year}{2006}).

\bibitem[{\citenamefont{Erban et~al.}(2007)\citenamefont{Erban, Frewen, Wang,
  Elston, Coifman, Nadler, and Kevrekidis}}]{frewenchem}
\bibinfo{author}{\bibfnamefont{R.}~\bibnamefont{Erban}},
  \bibinfo{author}{\bibfnamefont{T.}~\bibnamefont{Frewen}},
  \bibinfo{author}{\bibfnamefont{X.}~\bibnamefont{Wang}},
  \bibinfo{author}{\bibfnamefont{T.}~\bibnamefont{Elston}},
  \bibinfo{author}{\bibfnamefont{R.}~\bibnamefont{Coifman}},
  \bibinfo{author}{\bibfnamefont{B.}~\bibnamefont{Nadler}}, \bibnamefont{and}
  \bibinfo{author}{\bibfnamefont{I.~G.} \bibnamefont{Kevrekidis}},
  \bibinfo{journal}{Journal of Chem. Phys.} \textbf{\bibinfo{volume}{126}},
  \bibinfo{pages}{155103} (\bibinfo{year}{2007}).

\bibitem[{\citenamefont{Laing et~al.}(2007)\citenamefont{Laing, Frewen, and
  Kevrekidis}}]{laing}
\bibinfo{author}{\bibfnamefont{C.~R.} \bibnamefont{Laing}},
  \bibinfo{author}{\bibfnamefont{T.}~\bibnamefont{Frewen}}, \bibnamefont{and}
  \bibinfo{author}{\bibfnamefont{I.~G.} \bibnamefont{Kevrekidis}},
  \bibinfo{journal}{Nonlinearity} \textbf{\bibinfo{volume}{20}},
  \bibinfo{pages}{2127} (\bibinfo{year}{2007}).

\bibitem[{\citenamefont{Kevrekidis et~al.}(2003)\citenamefont{Kevrekidis, Gear,
  Hyman, Kevrekidis, Runborg, and Theodoropoulos}}]{man1}
\bibinfo{author}{\bibfnamefont{I.~G.} \bibnamefont{Kevrekidis}},
  \bibinfo{author}{\bibfnamefont{C.}~\bibnamefont{Gear}},
  \bibinfo{author}{\bibfnamefont{J.}~\bibnamefont{Hyman}},
  \bibinfo{author}{\bibfnamefont{P.}~\bibnamefont{Kevrekidis}},
  \bibinfo{author}{\bibfnamefont{O.}~\bibnamefont{Runborg}}, \bibnamefont{and}
  \bibinfo{author}{\bibfnamefont{C.}~\bibnamefont{Theodoropoulos}},
  \bibinfo{journal}{Commun. Math. Sci.} \textbf{\bibinfo{volume}{1}},
  \bibinfo{pages}{715} (\bibinfo{year}{2003}).

\bibitem[{\citenamefont{Kevrekidis et~al.}(2004)\citenamefont{Kevrekidis, Gear,
  and Hummer}}]{man2}
\bibinfo{author}{\bibfnamefont{I.~G.} \bibnamefont{Kevrekidis}},
  \bibinfo{author}{\bibfnamefont{C.}~\bibnamefont{Gear}}, \bibnamefont{and}
  \bibinfo{author}{\bibfnamefont{G.}~\bibnamefont{Hummer}},
  \bibinfo{journal}{AlChE Journal} \textbf{\bibinfo{volume}{50}},
  \bibinfo{pages}{1346} (\bibinfo{year}{2004}).

\bibitem[{\citenamefont{Kirkpatrick et~al.}(1983)\citenamefont{Kirkpatrick,
  Gelatt, and Vecchi}}]{sciencesimann}
\bibinfo{author}{\bibfnamefont{S.}~\bibnamefont{Kirkpatrick}},
  \bibinfo{author}{\bibfnamefont{C.~D.} \bibnamefont{Gelatt}},
  \bibnamefont{and} \bibinfo{author}{\bibfnamefont{M.~P.}
  \bibnamefont{Vecchi}}, \bibinfo{journal}{Science, New Series}
  \textbf{\bibinfo{volume}{220}}, \bibinfo{pages}{671} (\bibinfo{year}{1983}).

\bibitem[{\citenamefont{Nadler et~al.}(2006)\citenamefont{Nadler, Lafon,
  Coifman, and Kevrekidis}}]{coiffp}
\bibinfo{author}{\bibfnamefont{B.}~\bibnamefont{Nadler}},
  \bibinfo{author}{\bibfnamefont{S.}~\bibnamefont{Lafon}},
  \bibinfo{author}{\bibfnamefont{R.}~\bibnamefont{Coifman}}, \bibnamefont{and}
  \bibinfo{author}{\bibfnamefont{I.~G.} \bibnamefont{Kevrekidis}},
  \bibinfo{journal}{Advances in Neural Information Processing Systems}
  \textbf{\bibinfo{volume}{18}}, \bibinfo{pages}{955} (\bibinfo{year}{2006}).

\bibitem[{\citenamefont{Grassberger and Procaccia}(1983)}]{loglog}
\bibinfo{author}{\bibfnamefont{P.}~\bibnamefont{Grassberger}} \bibnamefont{and}
  \bibinfo{author}{\bibfnamefont{I.}~\bibnamefont{Procaccia}},
  \bibinfo{journal}{Physica D} \textbf{\bibinfo{volume}{9}},
  \bibinfo{pages}{189} (\bibinfo{year}{1983}).

\end{thebibliography}

\end{document}